\documentclass[aps,prb,reprint,superscriptaddress,longbibliography]{revtex4-2}
\usepackage{graphicx}
\usepackage{amsmath}
\usepackage{amssymb}
\usepackage{natbib}
\usepackage{color}
\usepackage{hyperref}

\begin{document}
\title{Record-large magnetically driven polarization in room temperature ferromagnets Os$X_2$ monolayers}
\author{Ying Zhou}
\author{Haoshen Ye}
\affiliation{Key Laboratory of Quantum Materials and Devices of Ministry of Education, School of Physics, Southeast University, Nanjing 211189, China}
\author{Junting Zhang}
\affiliation{School of Materials Science and Physics, China University of Mining and Technology, Xuzhou 221116, China}
\author{Shuai Dong}
\email{sdong@seu.edu.cn}
\affiliation{Key Laboratory of Quantum Materials and Devices of Ministry of Education, School of Physics, Southeast University, Nanjing 211189, China}

\begin{abstract}
Magnetically induced ferroelectrics in multiferroics provide an optimal approach to pursuit intrinsically strong magnetoelectricity. However, the complex antiferromagnetism, faint magnetically induced polarization, and low working temperatures make their magnetoelectric performance incompetent from the applications demands. Here, a family of two-dimensional $5d$ halides Os$X_2$ monolayers is predicted to be ferroelectric and ferromagnetic above room temperature. More interestingly, benefiting from the strong spin-orbital coupling and high-spin state of Os$^{2+}$ ion, the magnetically induced ferroelectric polarization can reach $5.9$ $\mu$C/cm$^2$, a record-large value in type-II multiferroics. The magnetoelectric effect, that is, controlling ferroelectric polarization by magnetic field has been demonstrated, and magnetically driven ferrovalley also emerges in this system. This work provides an effective way to solve the main defects of type-II multiferroics.  
\end{abstract}
\maketitle

\section{Introduction}
With the increasing demands for high-performance electronic devices, multiferroic materials with more than one ferroic order have attracted much attention \cite{Gong2017,Li2017,Dong2015,Dong2019,Spaldin2019,Spaldin2017}. The intrinsic coupling between/among these ferroic orders can lead to cross manipulations, which can be empolyed in various applications \cite{Zhong2017,Manchon2015}. Depending on the origin of ferroelectricity, multiferroics can be categorized into the type-I and type-II ones \cite{Khomskii2009}. The type-I multiferroics with independent origins of magnetism and ferroelectricity, are believed to be naturally weak regarding the intrinsic magnetoelectricity, even if their ferroelectricity and/or magnetism may be prominent \cite{Wang2003}. In contrast, the type-II multiferroics with polarization generated by particular magnetism, was highly desired, for their intrinsic strong magnetoelectricity \cite{Kimura2008,Tokunaga2009,Zhang2022}.

However, despite the great progress in the past decades, there remain some serious and inevitable drawbacks for type-II multiferroics, which hinder their further development. First, the magnetically induced ferroelectric polarization is very weak comparing with that of conventional proper ferroelectrics ($10-100$ $\mu$C/cm$^2$) \cite{Wang2003,Jeong2012,VanAken2004}. For example, for typical type-II multiferroics with spiral spin order, their polarization is $0.016$ $\mu$C/cm$^2$ and $0.08$ $\mu$C/cm$^2$ for CuO and TbMnO$_3$ \cite{Kimura2008,Kimura2003},  respectively. Relative larger polarization is derived from the nonrelativistic exchange-striction mechanism, which leads to $0.12$ $\mu$C/cm$^2$ and $0.8$ $\mu$C/cm$^2$ in the GdFeO$_3$ and $o$-YMnO$_3$ \cite{Tokunaga2009,Nakamura2011}, respectively. Second, most type-II multiferroics are antiferromagnets with complex spin textures (e.g., noncollinear/nonplanar, staggered/zigzag) originated from magnetic frustrations, which not only reduce the ordering temperatures but also lead to weak responses to external magnetic fields \cite{Zhang2018,Gong2019,Song2022}.

In another branch (the so-called linear magnetoelectricity), the magnetically induced polarization is linearly proportional to the magnetic field. However, the magnetically induced polarization is also typically faint because the underlying mechanisms are similar to those of type-II multiferroics\cite{Dong2019}. Even for the so-called ``colossal linear magnetoelectricity'' in Fe$_2$Mo$_3$O$_8$, its magnetically induced polarization is only $\sim0.6$ $\mu$C/cm$^2$ \cite{Chang2023}.

The $5d$ electron systems may be the solution to resolve this long-term predicament \cite{Hu2024}. The spin-orbital coupling (SOC), which usually plays as the glue between magnetism and polarity \cite{Dong2019}, is much larger in $5d$ electron systems. In addition, the broad spatial extension of $5d$ electron cloud, is helpful to enhance the orbital overlaps and thus the exchange interactions, a key factor to improve working temperature. Despite these positive factors, the known $5d$ multiferroics remain rare due to the following reasons. Their weak Hund coupling often fails when competing with the crystal field, resulting in a low spin state for $5d$ ions, which is disadvantageous to stabilize a local magnetic moment. In addition, the broad extension of $5d$ orbitals leads to a large bandwidth, which tends to overcome the weak Hubbard $U$ and result in a metallic state. 

\begin{figure}
\centering
\includegraphics[width=0.48\textwidth]{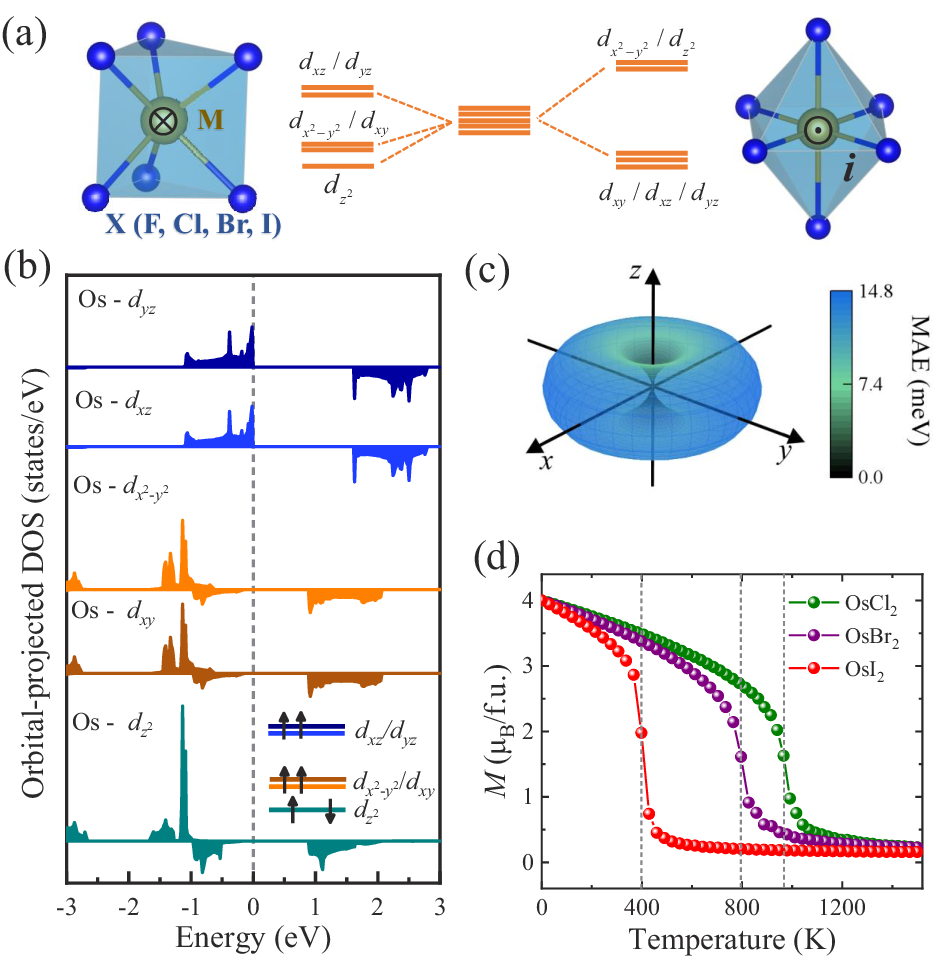}
\caption{(a) Schematic of the prism-type and octahedral crystalline field splittings of $d$-orbitals. The "$\bigotimes$" and "$\bigodot$" on magnetic ions represent the breaking and preservation of the inversion center ($i$), respectively. (b) The orbital-projected density of states (DOS) of  OsCl$_2$ monolayer. (c) The polar plot of MAE. (d) The MC simulations for ferromagnetic transitions of Os$X_2$ monolayers.}
\label{Fig1}
\end{figure}

Therefore, to pursuit type-II multiferroics containing $5d$ ions, the delicate balance among multiple interactions should be carefully tuned. First, the crystal field splitting should be minimized to obtain the high spin state. Comparing with divalent O$^{2-}$ and trivalent $N^{3-}$, the halogen $X^{-}$ is less charged, which can systematically reduce the crystal filed \cite{An2020}. Thus, halides are preferred over oxides and nitrides. Second, the bandwidth of $5d$ orbitals should be moderate to avoid metallicity. Two-dimensional (2D) van der Waals materials may be proper candidates, in which the quantum confinement effect can narrow the bandwidth and increase the band gap to a certain extent. For 2D metal halides, there are two common families $MX_2$ and $MX_3$ monolayers, which form the triangular prism and octahedral coordination, respectively. As shown in Fig.~\ref{Fig1}(a), the former has a relatively smaller splitting gaps. More importantly, the octahedral coordination preserves the inversion center, thus ruling out the possibility of magnetically induced ferroelectricity in ferromagnets. Therefore, the $MX_2$ monolayer systems are promising candidates for type-II multiferroics with excellent magnetoelectricity.

In this paper, following the above design principles, we predict Os$X_2$ monolayers as $5d$ type-II multiferroics with prominent performance by using first-principles calculation. Our results show that in this system, magnetically induced ferroelectricity and ferromagnetism are compatible and coupled, with Curie temperature above room temperature. The dependence of the induced polarization on the spin direction can be explained by the $p-d$ hybridization mechanism involving SOC. The strong SOC of Os ions leads to a large polarization of $5.9$ $\mu$C/cm$^2$, the largest value among known type-II multiferroic materials. Besides ferroelectricity and ferromagnetism, magnetically driven ferrovalley also emerges and can be controlled by magnetic field. 

\section{Computational Methods}
The first-principles calculations based on density functional theory (DFT) were performed using the projector-augmented wave method, as implemented in the Vienna \emph{ab initio} Simulation Package \cite{Kresse1996,Kresse19961,BLOCHL1994}. The Perdew-Burke-Ernzerhof functional was used as the exchange-correlation functional \cite{Kresse1999,WANG1991}. A vacuum space of $20$ {\AA} was added to avoid interaction between neighboring periodic images. We used a cutoff energy of $600$ eV for the plane-wave bases, and a $\varGamma$-centered $15\times15\times1$ $k$-point mesh for the Brillouin zone integration. The in-plane lattice constants and internal atomic coordinates of each structural phase were relaxed until the Hellman-Feynman force on each atom is less than $0.005$ eV/{\AA}. A convergence threshold of $10^{-7}$ eV was used for the electronic self-consistency loop.

To describe correlated $5d$ electrons of Os, the GGA+$U$ method is applied \cite{Dudarev1998}, and the $U_{\rm eff}$ is tested from $0$ to $2$ eV [see Supplementary Materials (SM) for more details \cite{SM}]. For comparison, Heyd-Scuseria-Ernzerhof (HSE06) hybrid functional was also considered \cite{Heyd2003}. Since the band structure at $U_{\rm eff}=1.5$ eV is the most consistent one with that obtained from the HSE06 hybrid functional, $U_{\rm eff}=1.5$ eV will be adopted as default in the following discussion.

Phonon band structures were calculated using density functional perturbation theory (DFPT). The phonon frequencies and corresponding eigen-modes were calculated on the basis of the extracted force-constant matrices with a $4\times4\times1$ supercell, as implemented in the PHONOPY code \cite{Togo2015}. The ferroelectric polarization was calculated by using the Berry phase method \cite{KINGSMITH1993}.

In addition, Monte Carlo simulations were performed to verify the magnetic ground states and estimate the magnetic transition temperatures.
The $T_{\rm C}$ is estimated via the Monte Carlo (MC) simulation on the classical spin model, which can be written as:
\begin{equation}
	H=-J_{ij}\sum_{\langle i,j\rangle} {{{\bf{S}}_i} \cdot {{\bf{S}}_j}}  - A\sum\limits_i {{{({\bf{S}}_i^z)}^2}},
	\label{ham}
\end{equation}
with $J$ and $A$ as the exchange parameters and the anisotropy constant. The nearest neighboring $J_1$, next-nearest neighoring $J_2$ , and third-nearest neighoring $J_3$ are considered, which can be extracted from DFT energies. A $30\times30$ triangular lattice is used in our MC simulation. MC steps for both thermal equilibrium and measurement sampling by Metropolis algorithm are $10000$ steps.

\section{Results and discussion}
\subsection{Crystal structure and magnetism}
Figure~\ref{Fig1}(a) illustrates the crystal structure of Os$X_2$ ($X$= F, Cl, Br, and I) monolayers with space group $P\overline{6}m2$ (the 2H-phase of $MX_2$), which is nonpolar and noncentrosymmetric. As shown in Fig. S1 in SM \cite{SM}, there is no imaginary modes in phonon dispersions of all Os$X_2$ except OsF$_2$, indicating dynamic stability of Os$X_2$ ($X$= Cl, Br, and I). This provides a solid guide for successful experimental fabrication. Like other 2D ferromagnetic materials such as VSe$_2$ \cite{Yu2019,Manuel2018} and CrI$_3$ \cite{Huang2017}, these  Os$X_2$ might be grown using methods like chemical vapor deposition (CVD) or chemical vapor transport (CVT), followed by mechanical exfoliation down to monolayers. Subsequently, characterization techniques such as atomic force microscopy (AFM), scanning electron microscopy (SEM), and X-ray photoelectron spectroscopy (XPS) are employed to analyze their structure and properties. As these three monolayers share the similar properties, we only focus on monolayer OsCl$_2$ as a representative in the following.

The optimized lattice constant of OsCl$_2$ monolayer is $3.43$ {\AA}. For every Os$^{2+}$ ion, the six unpaired electrons occupy the $5d$ orbitals and form the high spin state, i.e., five spin-up and one spin-down mainly in the $3z^2-r^2$ orbital, as shown in Fig.~\ref{Fig1}(b). Such high spin state can only be stablized when the crystal filed spliting is weak between $xz$/$yz$ and $x^2-y^2$/$xy$ orbitals. Consequently, the Os$^{2+}$ exhibits a large magnetic moment $4$ $\mu_{\rm B}$, which is confirmed in our DFT calculation. The magnetocrystalline anisotropy energy (MAE) is calculated with SOC enabled as the energy difference between different spin directions. Our result indicates an easy magnetic axis along the out-of-plane direction for the OsCl$_2$ monolayer, as shown in Fig.~\ref{Fig1}(c). Due to the strong SOC, its MAE is rather large, reaching $14.8$ meV/Os.

Furthermore, by comparing the ferromagnetic and three most possible anti-ferromagnetic configurations (as shown in Fig. S2 in SM \cite{SM}), the ferromagnetic state is found to be the most energetically stable, as summarized in Table S1 in SM \cite{SM}. Using the nomarlized spin $|S|=1$, the nearest-neighbour exchange $J_1$ is estimated as $77.1$ meV, a quite strong ferromagnetic coupling, while the second- and third-neighbour ones are relative weaker ($J_2=10.0$ meV and $J_3=-12.0$ meV). Then the ferromagnetic Curie temperature $T_{\rm C}$ of OsCl$_2$ monolayer is predicted to be $\sim966$ K via MC simulation of class spin model, as shown in Fig.~\ref{Fig1}(d).

The magnetic properties of OsBr$_2$ and OsI$_2$ are also calculated, as compared in Table~\ref{tab1}. Both of them are ferromagnetic with a common magnetic easy axis pointing out-of-plane (i.e. $z$-axis). It can be found that the $J_1$ and the associated $T_{\rm C}$ are much larger/higher in OsCl$_2$ than other two. The physical reason is the much shorter Os-Cl bond ($2.61$ \AA) than others ($2.74$ \AA{} for Os-Br and $2.91$ \AA{} for Os-I) due to the smallest size and strongest electronegativity of Cl$^-$ ion, which strengths the $p-d$ orbital hybridization. Even though, the estimated $T_{\rm C}$ of OsBr$_2$ and OsI$_2$ remain above room temperature. It is mainly due to the large ferromagnetic $J_1$, which are natural superiority of $5d$ magnets, inherited from its more spatial expansion of electronic clouds. In addition, the MAE increases with atomic number of $X$, as expected from the increasing SOC.

For completeness, the magnetism of of Os$X_2$ are tested as a function of $U_{\rm eff}$ from $0$ to $2$ eV, as summarized Tables S1-S4 in SM \cite{SM}. For the OsCl$_2$ and OsBr$_2$, the MAE monotonically decreases as $U_{\rm eff}$ increases from $0$ to $2$ eV. It can be understood from second-order perturbation theory analysis \cite{Wang1993},which related to the valence and conduction bands move further away from the Fermi level as $U_{\rm eff}$ increases, as shown in Fig. S3 in SM \cite{SM}. For most $U_{\rm eff}$, there are no qualitative differences, except for $U_{\rm eff}=0$ eV for OsI$_2$.

\subsection{Ferroelectricity}

\begin{table}
	\centering
	\caption{Ferromagnetic, ferroelectric, and ferrovalley properties of Os$X_2$ monolayers with $U_{\rm eff}=1.5$ eV. The MAE ($A$), exchange coefficients ($J$'s), magnetic moment ($M$), band gap, the maximum in-plane and out-of-plane polarization ($P_x^{max}$, $P_z^{max}$) are presented. $J_1$,  $J_2$, and $J_3$ denote the nearest-neighbor, next-nearest-neighbor, and third-nearest-neighbor exchange interactions, respectively. To calculate the polarization, the thickness of monolayers are used to estimate the volumes.}
	\begin{tabular*}{1.0\linewidth}{@{\extracolsep{\fill}}lccc}
		\toprule
		Materials & OsCl$_2$ & OsBr$_2$ & OsI$_2$  \\
		\hline
		$A$ (meV/f.u.) & 14.8  &16.5 & 30.0  \\
		$J_1$ (meV/f.u.) & 77.1 & 56.8 & 35.4  \\
		$J_2$ (meV/f.u.) & 10.0  & 3.1 & -0.1  \\
		$J_3$ (meV/f.u.) & -12.0  & -3.8 & -6.7  \\
		$T_C$ (K)       & 966  & 796 & 398 \\
		$M$ ($\mu_{\rm B}$/Os) & 4.0   & 4.0 & 4.0  \\
		Band gap (meV) & 470   & 415 & 21  \\
		$P_x^{max}$ ($\mu$C/cm$^2$) & 2.5   & 2.6 & 5.9  \\
		$P_z^{max}$ (nC/cm$^2$) & 6.9   & 15.5 & 31.2  \\
		$\Delta E_K^{\rm VBM}$ (meV) & 128   & 77 & 47 \\
		\hline
	\end{tabular*}
	\label{tab1}
\end{table}%

\begin{figure}
	\centering
	\includegraphics[width=0.48\textwidth]{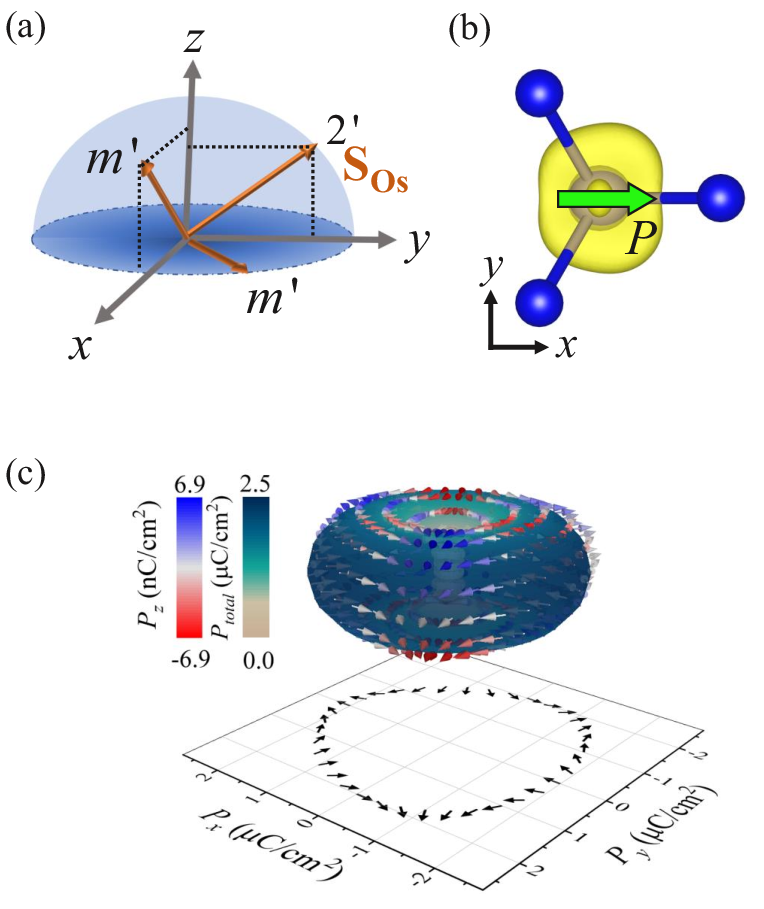}
	\caption{(a) The evolution of MPG as the spin rotates within the $xy$ plane, $xz$ plane, and $yz$ plane. The orange arrow denotes the spin of Os ($\textbf{S}_{\rm Os}$). Here $x$-axis and $z$-axis are in-plane 2-fold rotation axis and out-of-plane 3-fold rotation axis, respectively. And the $xz$ plane and $xy$ plane are vertical mirror plane and horizontal mirror plane, respectively. The $m'$ within the xz plane preserves thevertical mirror plane, while the $m'$ within the xy plane preserves the horizontal mirror plane. (b) Differential electron density between the states with spin along the $x$-axis and $z$-axis. (c) Three-dimensional polar plot of polarization for OsCl$_2$. The projection to the $xy$ plane is also shown. For better stereoscopic view, the value of $P_z$ component is magnified for $300$ times.}
	\label{Fig2}
\end{figure}

We next investigated the ferroelectricity of the Os$X_2$ monolayers. The magnetic point group (MPG) for this ground state (out-of-plane magnetocrystalline anisotropy) is $\overline{6}m'2'$, which is nonpolar. Polarization can be induced when spin rotates away from the $z$-axis, as shown in Fig.~\ref{Fig2}(a). For spin along arbitrary direction in the $xz$ plane (except the $x$- and $z$-axes), the in-plane 2-fold rotation symmetry and horizontal mirror symmetry are broken, and thus the MPG becomes $m'$, as shown in Fig.~\ref{Fig2}(a). Thus, both the in-plane and out-of-plane components of polarization (i.e. $P_x$ \& $P_z$, and $P_y=0$) are allowed. The corresponding polar plots of $P_x$ and $P_z$ are shown in Fig. S4(b) in SM \cite{SM}. The numerical values obtained via DFT calculation agree with the analytic formula very well. The maximum value of the in-plane polarization component reach $2.5$ $\mu$C/cm$^2$ for OsCl$_2$ monolayer, while the value of the out-of-plane polarization component is only $6.9$ nC/cm$^2$.

This magnetically induced polarization can be visualized using the differential electron density between the two states: spin along the $x$- and $z$-axis, as shown in Fig.~\ref{Fig2}(b). The in-plane charge distribution along the ligand directions is not uniform, leading to the appearance of in-plane polarization.

For spin along arbitrary direction in the $yz$ plane (except the $y$- and $z$-axes), the horizontal and vertical mirror symmetries are broken, and thus the MPG becomes $2'$, and only the in-plane $P_x$ component of polarization is allowed. The corresponding polar plot of $P_x$ is shown in Fig. S4(c) in SM \cite{SM}. For spin along other arbitrary direction (except in the $xy$/$xz$/$yz$ planes), all symmetries are broken, and MPG becomes $1$. Then all components of polarization are allowed. A three-dimensional polar plot (and its projection to the $xy$ plane) of $\textbf{P}$ obtained in DFT calculation is shown Fig.~\ref{Fig2}(c), which forms a horn torus.

The origin of polarization can be explained by the $p-d$ hybrid mechanism which related to the SOC effect \cite{Murakawa2010}. It can be obtained by summing the charge transfers between the magnetic ion and its ligands, expressed as $\textbf{P} \propto \sum_{i}^n(\textbf{S}\cdot \textbf{e}_{i})^2\cdot\textbf{e}_{i}$, where \textbf{e}$_{i}$ is the unit vector of the bond between the magnetic ion and ligands, and \textbf{S} is the vector of spin. Therefore, the net $\textbf{P}$ of Os$X_2$ monolayers can be expressed as:
\begin{equation}
	\textbf{P}=(P_x, P_y, P_z)\propto(S_x^2 - S_y^2, -2S_xS_y, 0),
\label{P}
\end{equation}
which can also describe the magnetically induced polarization in other 2H-type ferromagnetic monolayers such as VSe$_2$, VTe$_2$, and RuBr$_2$ \cite{Wang:PRA,Zhou2024}. However, Eq.~\ref{P} can only explain the major in-plane polarization, while the tiny out-of-plane $P_z$ obtained in DFT calculation must come from the high-order terms of the SOC effect \cite{Jia2006,Wang:PRA}. The out-of-plane polarization becomes:
\begin{equation}
	P_z\propto(4S_x^3S_z+4\sqrt{3}S_x^2S_yS_z+12S_xS_y^2S_z).
	\label{Pz}
\end{equation}
Since here the magnetically induced polarization is generated via a SOC process, the intensity of SOC plays the key role to the magnitude of polarization. It is the reason why OsCl$_2$ owns a much larger polarization than VSe$_2$ \cite{Wang:PRA} due to the strong SOC of $5d$ orbitals. Furthermore, the polarization of OsBr$_2$ and OsI$_2$ can be even larger due to the additional large SOC from anions, as compared in Table~\ref{tab1}. The largest polarization reaches $\sim5.9$ $\mu$C/cm$^2$ in OsI$_2$ monolayer ($U_{\rm eff}=1.5$ eV), a record-large value in magnetically induced polarization and even comparable to the magnitude of geometric ferroelectrics \cite{Benedek2011}. Meanwhile, the out-of-plane $P_z$ is also enhanced by larger SOC from anions, as shown in Table~\ref{tab1}.

\begin{figure}
	\centering
	\includegraphics[width=0.48\textwidth]{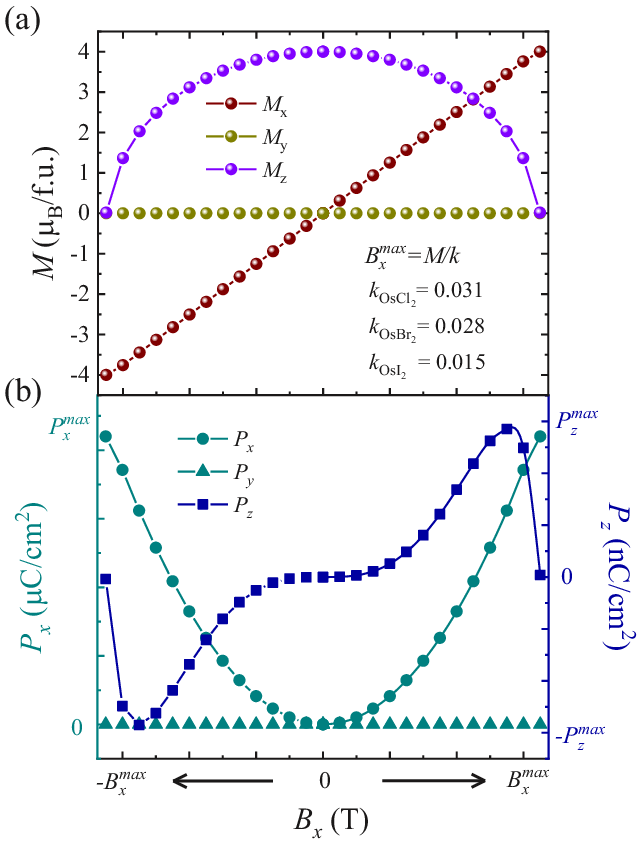}
	\caption{Magnetic field tuning of magnetoelectric properties of Os$X_2$ monolayers at zero-temperature. The magnetic field is applied along the $x$-axis. (a) Three components of magnetization. The $x$-component of magnetization is in proportional to the field linearly. (b) Three components of polarization components. The $x$-component of polarization is in a parabolic function of field. $B_x^{max}$ is the field required for magnetizatic saturation.}
	\label{Fig3}
\end{figure}

The strong coupling between spin and polarization makes magnetic field an effective strategy to manipulate the polarization of Os$X_2$ monolayers. By considering the MAE and Zeeman energy, a model Hamiltonian can be expressed as:
\begin{equation}
	E = E_{\rm MAE} + E_{B_x}= A\cdot\sin^2\theta - M\sin\theta B_x,
\end{equation}
in which $B_x$ denotes the external magnetic field along the $x$-axis, $M$ is the magnetization $4$ $\mu_{\rm B}$/Os, and $\theta$ is the polar angle. The the equilibrium state can be obtained by $\partial E/\partial\theta=0$. Then it is straightforward to obtain the relationship between the in-plane component of magnetization ($M_x$) and $B_x$ as:
\begin{equation}
	M_x = 2MB_x/A,
\end{equation}
which is a linear function with a slop coefficient $k=2M/A$ before the saturation, as shown in Fig.~\ref{Fig3}(a).

Meanwhile, the magnetically induced in-plane polarization $P_x\propto S_x^2\propto M_x^2$. Therefore, $P_x\propto B_x^2$ [Fig.~\ref{Fig3}(b)], a parabolic behavior different from the well known linear magnetoelectricity.

\subsection{Ferrovalley}
\begin{figure}
\centering
\includegraphics[width=0.48\textwidth]{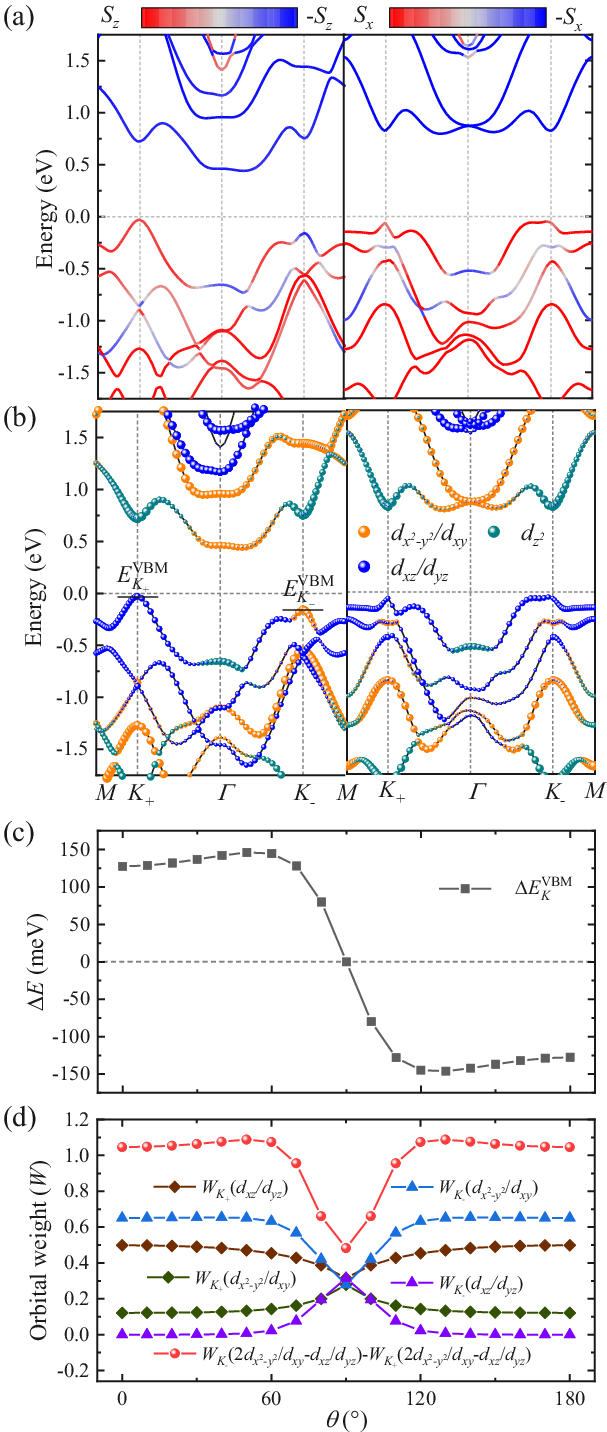}
\caption{Ferrovalley of OsCl$_2$ monolayer. (a) Spin-projected band structures and (b) Orbital-projected band structures obtained at $U_{\rm eff}=1.5$ eV. Left panels: spin points along $z$-axis; Right panels: spin points along $x$-axis.(c) The magnitude of valley splitting of the VBM at the $K_{+/-}$ points as a function of the spin polar angle. (d) The orbital weights of the VBM at the $K_{+/-}$ points as a function of the spin polar angle.}
\label{Fig4}
\end{figure}

In addition to ferromagnetism and ferroelectricity, Os$X_2$ monolayers can also exhibit the ferrovalley degree of freedom. Figure~\ref{Fig4} shows their band structures at $U_{\rm eff}=1.5$ eV. The VBM are different at the $K_+$ and $K_-$ points when spin points along the $z$-axis, as shown in the left panels of Fig.~\ref{Fig4}. Here the valley polarization is defined as $\Delta E_K=E_{K_+}-E_{K_-}$, where $E_{K_+}$ and $E_{K_-}$ refer to the energies of valence band maximum (VBM) at the $K_+$ and $K_-$ points, respectively. For OsCl$_2$ monolayer, $\Delta{E}_K^{\rm VBM}$ reaches $128$ meV. This valley polarization disappears when spin lies in the $xy$-plane as shown in the right panels of Fig.~\ref{Fig4}. The valley polarization of OsBr$_2$ and OsI$_2$ monolayers are summarized in Table~\ref{tab1}.

Although the ferrovalley properties generally exist in many monolayers, here OsCl$_2$ exhibits some unique phenomena. In other ferrovalley materials such as VSe$_2$ and RuCl$_2$, neither the spin nor the orbital projection of the VBM undergoes significant change when the valley polarization is switched \cite{Cui2021,Sheng2022,Peng2020}. In contrast, here in the ferrovalley state with spin pointing along the $z$-axis, the dominant spin component of VBM at the $K_-$ valley is reversed [Fig.~\ref{Fig4}(a)], and the dominant orbital component of VBM is also changed at the $K_-$ valley [Fig.~\ref{Fig4}(b)].

Such significant spin-orbital reconstruction of valley state is a consequence of strong SOC of Os ion \cite{Weng:PRB,Weng:PRB1}, which will lead to nontrivial tuning of ferrovalley polarization. Figure~\ref{Fig4}(c) shows $\Delta{E}_K$'s of VBM as a function of spin polar angle $\theta$. The VBM curve is anomalous, which reaches the maximum value when $\theta$ is $60^\circ$ or $120^\circ$. For comparison, the largest valley polarization always appears when spin points along the $z$-axis in VSe$_2$ \cite{Wang:PRA}, VSi$_2$N$_4$ \cite{Cui2021}, and Nb$_3$I$_8$ \cite{Peng2020}.

The orbital weights ($W$'s) of the VBM obtained in DFT calculations may provide an explanation for the aforementioned phenomenon. As show in Fig.~\ref{Fig4}(d), when spin points along the $z$-axis ($\theta=0^\circ$), the $K_-$ valley is primarily contributed by the $d_{x^2-y^2}$/$d_{xy}$ orbitals, while the $K_+$ valley is primarily contributed by the $d_{xz}$/$d_{yz}$. When the spin pointing along the $x$-axis ($\theta=90^\circ$), the contributions from $d_{xz}$/$d_{yz}$ orbitals and $d_{x^2-y^2}$/$d_{xy}$ orbitals are almost equal to the VBM's of both $K_+$ and $K_-$ valleys. Interestingly, the sum of orbital weights differences of $K_+$ and $K_-$ also exhibits the maximum value when $\theta$ is $60^\circ$ or $120^\circ$.

\section{Conclusion}
In summary, our work has demonstrated that Os$X_2$ monolayers are candidate low-dimensional functional materials exhibiting excellent ferromagnetic, ferroelectric, and ferrovalley properties, as well as magnetoelectric and magneto-valley coupling effects. In particular, Os$X_2$ monolayers own high ferromagnetic Curie temperatures above room temperature, and their magnetically induced polarization can reach several $\mu$C/cm$^2$. A parabolic magnetoelectric behavior is predicted, distinguishing them from those linear magnetoelectricity. Furthermore, the ferromagnetic ground state protects the intrinsic valley polarization.

\begin{acknowledgments}
This work was supported by the National Natural Science Foundation of China (Grant Nos. 12325401, 12274069, \& 12374097) and the Big Data Computing Center of Southeast University.
\end{acknowledgments}

\bibliography{reference}

\end{document}